# On the differences between citations and altmetrics: An investigation of factors driving altmetrics vs. citations for Finnish articles[1]

**Fereshteh Didegah (Corresponding author)[1], Timothy D. Bowman, & Kim Holmberg**

*Research Unit for the Sociology of Education*
*University of Turku, Finland*

[1]*Fereshteh.didegah@utu.fi*

**Abstract:** This study examines a range of factors associating with future citation and altmetric counts to a paper. The factors include journal impact factor, individual collaboration, international collaboration, institution prestige, country prestige, research funding, abstract readability, abstract length, title length, number of cited references, field size, and field type and will be modelled in association with citation counts, Mendeley readers, Twitter posts, Facebook posts, blog posts, and news posts. The results demonstrate that eight factors are important for increased citation counts, seven different factors are important for increased Mendeley readers, eight factors are important for increased Twitter posts, three factors are important for increased Facebook posts, six factors are important for increased blog posts, and five factors are important for increased news posts. Journal impact factor and international collaboration are the two factors that significantly associate with increased citation counts and with all altmetric scores. Moreover, it seems that the factors driving Mendeley readership are similar to those driving citation counts. However, the altmetric events differ from each other in terms of a small number of factors; for instance, institution prestige and country prestige associate with increased Mendeley readers and blog and news posts, but it is an insignificant factor for Twitter and Facebook posts. The findings contribute to the continued development of theoretical models and methodological developments associated with capturing, interpreting, and understanding altmetric events.

# Introduction

Traditionally, citations have been used as the main indicators for measuring research impact (Bornmann & Daniel, 2008), but they have been criticized for not being able to reflect a broader impact of research (Holmberg, Didegah, & Bowman, 2015), such as educational, cultural, environmental, and economic impact. For this purpose, some novel data sources and indicators are being investigated. The term 'altmetrics' is described as the collection and measure of online events relating to the consumption and dissemination of scientific documents and the filtering of scientific documents from the vast stream of information that is available online (Priem, Taraborelli, Groth, & Neylon, 2010). Altmetrics seeks to investigate mentions of various research outputs, such as scientific articles and dataset, from online sources with the assumption that these mentions could reveal something about the impact or influence research has had on different audiences beyond academia (Priem, 2014). Earlier altmetrics research focused on investigating possible connections between mentions of scientific articles on different online platforms and the number of citation those articles later received, finding evidence of such a connection ofr instance for articles that were bookmarked on Mendeley (Thelwall & Wilson, 2016), mentioned in Wikipedia (Evans & Krauthammer, 2011) and tweeted on Twitter (e.g. Thelwall et al., 2013). More recently altmetrics research has moved on to incorporate other approaches in its pursue to explain the meaning of and motivations behind altmetrics (e.g., Vainio & Holmberg, 2017).

---

[1] This is a pre-print of article accepted for publication in JASIS&T, copyright 2017.

As motivations to create citations can vary, a large number of studies (see Didegah (2014) for a comprehensive review of factors examined) have investigated the reasons and motivations for creating citations in order to help scholars and institutions better understand what attracts citations to scientific papers, but there are very few studies trying to investigate factors and reasons for mentions of research articles on altmetric platforms. A survey of 679 Mendeley users showed that the main motivation for adding articles to Mendeley library was to cite them later (Mohammadi, 2014). Another research on Twitter showed that articles are mainly posted on Twitter by scholars themselves for publicizing purposes (Thelwall, Tsou, Weingart, Holmberg, & Haustein, 2013). Articles with funny and light titles and common social topics were also more tweeted (Neylons, 2014; Didegah, Bowman, Bowman, & Hartley, 2016).

This study aims to measure some other factors and reasons in association with altmetric counts to research articles. These factors are mainly the characteristics of articles including: the impact of journal that the article is published in; whether the article is single author or multi-author or it is in collaboration with international peers; the prestige of institution and country that the author(s) of the article come from; whether the article is published from a funded project; how readable and long the article abstract is; how long the article title is; how many references the article has cited; and how large and what type the subject field of article is. These article characteristics have been widely examined in association with article traditional citation counts in previous studies. Some factors such as journal impact or number of authors were found to be important for citations but some other features such as field size were not significant factors (Didegah & Thelwall, 2013; Vanclay, 2013). So, it is interesting to measure whether important factors for citations can also be determinants of altmetric counts to articles.

Furthermore, altmetrics is also yet lacking a specific set of theories and frameworks to define its functions and applications. Haustein, Bowman and Costas (2015) made an attempt to apply existed citation and social theories to different altmetric platforms. Normative and social constructivist theories are among the most important citation theories that they theoretically discussed in relation to altmetric events. According to the social constructivist view of citations, research properties (i.e. article characteristics such as the ones mentioned above) other than research quality (that is the focus of normative theory of citations) can contribute to citation impact (White, 2004; Baldi, 1998; Gilbert, 1977). In addition, citations may occur for persuasion, perfunctory or critical reasons or a Matthew effect may cause increasing citations. Haustein, Bowman and Costas (2015) tried to theoretically apply this view to some altmetric events such as Mendeley and Twitter. For example, a social constructivist theory could be applied to Mendeley in relation with Mohammadi's (2014) study that found readers mainly save articles to cite them later in their own research; or for instance, they argue that there may be some forms of persuasion in posting articles on Twitter, or that there are negative mentions of articles in tweets.

Examining article characteristics in association with altmetric counts is actually an empirical attempt to test the social constructivist theory on altmetric events and the results can contribute to the development of a theory for altmetrics. Moreover, the results will lead to a better understanding of altmetrics and contribute to the discourse concerning the rationale behind the generation of such indicators and whether they can be considered as alternatives or complements to citations.

## Research background

Publishing in a high impact journal is an important signal for increasing attention to a research paper. Most studies confirm that journal impact factor is the most significant determinant of citations (Vanclay, 2013), however at least one study demonstrates an exception (Stremersch, Verniers & Verhoef, 2007); the similarity between the journals and the small sample size of this study may have affected the results for journal impact factor, as only the five top journals in marketing were taken into account.

Multi-author research has been widely found to have a citation advantage (Chen, 2012; Gazni & Didegah, 2010; Sooryamoorthy, 2009; Rousseau, 1992). Moreover, the number of authors is a significant factor in all subject fields, although the extent to which it associates with increased citations varies from 1.2% in Space Sciences to 16.3% in Economics and Business (Didegah, 2014). However, a few studies have found no correlation between additional authors and increased citations (Bornmann, Schier, Marx, & Daniel, 2012; Haslam et al., 2008). International collaboration is another important factor contributing to increased citation counts (Sin, 2011; Persson, 2010). While multi-institutional research has been found to receive more citations than single-institutional research (Sooryamoorthy, 2009), modeling this factor simultaneously with the other two above-mentioned patterns of collaboration demonstrated that it is not an important citation factor (Didegah, 2014).

Other citation factors have revealed that researchers from high-ranked institutions receive more citations to their papers than those from low-ranked institutions (Leimu & Koricheva, 2005), presumably (at least partly) because they tend to be better researchers. In addition, researchers from a particular country may produce papers with relatively higher impact (West & McIlwaine, 2002). For example, in ecological journals, UK authors receive more citations than authors from other European countries (Leimu & Koricheva, 2005).

Medical papers with longer abstracts have been found to receive more citations (Kostoff, 2007), whereas papers with longer titles in psychology seem to receive fewer citations (Haslam et al., 2008). Articles with a higher number of references also tend be cited more often (Vieira & Gomes, 2010).

Articles with more readable (easier) abstracts receive higher numbers of citations in Social Sciences only, while it is the opposite in Physical and Natural Sciences (Didegah, 2014).

Funded research in medical education research (Reed et al., 2007), library and information science (Zhao, 2010), and biomedical research (Lewison & Dawson, 1998) is more cited than unfunded research, although it may vary across subject domains in a single country (Jowkar, Didegah, & Gazni, 2011).

Field type, in terms of Natural Sciences versus Social Sciences or theoretical sciences versus applied sciences, is also a driver of citations (Kulkarni, Busse, & Shams, 2007; Callaham, Wears. & Weber, 2002; Peters & Van Raan, 1994), with natural and applied sciences having an advantage over the others. This advantage is also demonstrated in the UK

Research Assessment Exercise (RAE) from 2001, where the mean citation counts for biomedical articles was about 30, for social science articles 5, and for humanities articles 2 (Mahdi, D'Este, & Neely, 2008). Furthermore, articles in smaller fields normally receive fewer citations than those in more general fields (King, 1987) and for this reason the citation assessment of institutions is always related to the average citation impact of the field (Van Raan, 2003). In addition, open access journals are found to receive a higher number of citations than non-open access journals (Vanclay, 2013; Eysenbach, 2006).

As reviewed above, the factors included in the current study have been widely studied in association with citation counts in prior research, while studies into citation factors in the context of altmetrics are scarce. Haustein, Costas, and Larivière (2015) studied the association between discipline and document type, title, and paper length, number of references, and research collaboration with both citation and altmetric counts including blogs, Twitter, Facebook, Google+, mainstream media, and newspaper mentions. They concluded that factors driving citations and altmetric counts mostly differ from each other, although research collaboration and number of listed references in the research articles were found to increase both citation and altmetric counts. In another research, study types of medical articles that have been mentioned on Twitter was examined (Andersen & Haustein, 2015). Meta-analyses, systematic reviews and clinical trials were the most frequently tweeted types while basic research is the least frequently tweeted. The authors conclude that those study types that are more interesting to layman and patients have been tweeted more often. Didegah, Bowman, Bowman, and Hartley (2016) focused on the characteristics of titles of highly tweeted articles with that of highly mentioned articles on other platforms and highly cited articles and reported significant differences between the platforms such as that while highly tweeted articles may have funny and more general titles, highly cited articles in the Web of Science and also in Wikipedia had serious titles with more technical words. So, it would be interesting to investigate and compare more factors across different altmetric and citation platforms.

Using a more advanced simultaneous statistical model—a negative binomial-logit hurdle model—this research will study the association between twelve factors and six platforms. Previous studies have used mostly simple regression or correlation tests that do not allow a simultaneous assessment of factors. This is a key omission because inappropriate models may generate misleading conclusions and non-simultaneous tests may identify significant factors that are not relevant when other factors are also considered (Didegah, 2014).

## Research questions

While not the first to research altmetric factors, this work examines newer and additional factors contributing to altmetric events using an advanced statistical model. The goals of this study can be summarized in the following research questions:

1. Do the factors driving citations differ from those driving altmetric events?
2. Does the influence of the factors differ between each altmetric platform?

The research questions address the differences between citation and altmetric factors and attempt to determine whether or not similar factors are influencing both citations and altmetric indicators and whether or not altmetrics can be considered as alternatives or complements to citations. An increase in understanding how different factors affect altmetrics will allow for a greater understanding of the events and for a more complete definition of altmetrics. Based on the known similarities and differences between the platforms, such as the knowledge that scholars are primarily involved in citation databases and Mendeley, while both scholars and the general public are contributing to social media platforms, then the research hypothesis is that citation and Mendeley readership do behave similarly but there are differences between these two indicators with social media indicators.

**Methods**

*Sample:* All available records for documents with at least one author with an affiliation to a Finnish organization and published from 2012 to 2014 as indexed by the Web of Science (WoS), Thomson Reuters, were collected. To obtain the sample, a query was run in WoS by selecting Finland as the country and publishing years from 2012 to 2014. The time period of 2012-2014 was chosen as this is the timeframe during which online mentions have been more systematically identified and collected by Altmetric.com. We acknowledge that this may have an impact on the results whereas citations are concerned, as the citation window is limited. Thus, the results on the analysis of the association between citation counts and the other factors may rather be an underestimation of the actual situation. A total of 46,730 documents of all types and covering any disciplines were retrieved, out of which 37,183 documents were identified as having a DOI (Document Object Identifier). To obtain the number of tweets, Facebook posts, and blog and news posts, those documents that had a DOI were searched and matched with altmetric data (as collected by Altmetric.com) resulting in a final sample of 13,623 documents having altmetric events. Altmetric.com also retrieves Mendeley readership, but it only collects number of readers for DOIs that have already been tweeted, posted on Facebook, or mentioned in a news, etc. Therefore, the number of Mendeley readers was retrieved through Mendeley API separately; out of the 37,183 documents with DOIs, 35,980 documents were found in Mendeley and 35,972 documents had at least one reader (eight documents were added to the Mendeley library but had zero readers). As this study is comparing the factors associating with both 'non-zero altmetric events' and 'zero altmetric events', the entire Finnish dataset of 46,730 documents was considered for modeling; altmetric events (the number of tweets, and number of Facebook, blog, and news posts) and the number of readers for each document that were not found in Altmetric.com and/or Mendeley were assigned a zero value. Given that only 13,623 (30%) documents were matched in the Altmetric.com dataset, there was an excessive number of zeroes in the Twitter, Facebook, blog and news data. Therefore, the statistical model implemented in this study (as explained below) was chosen because it is designed to deal with excess zeroes. Hence, the results should not be biased by the number of zeroes in the datasets. A 30% match on altmetric.com may not represent the population especially that altmetric indicators have bias towards specific fields: for instance, Medical and Health research is more tweeted than other type of research. But this low match is due to limitations of altmetric tools that only retrieve documents using their DOIs and not any other bibliographic information which may affect the results of this study.

*Variables and measures:* There are two groups of dependent and independent variables in this study. Dependent variables include citation counts, Mendeley readers, Twitter posts, Facebook posts, blog posts, and news posts. Independent variables include journal impact factor, individual and international collaborations, institution and country prestige, research funding, abstract readability and length, title length, number of references, field size, and field type. It is difficult to analyze collinear variables since their effect on the outcome may result from either true associations or spurious correlations. Hence, as institutional collaboration is highly correlated with individual and international collaborations, this factor was not included in the model. In addition to the dependent variables listed above, open access was also taken into consideration and modeled by using data from the Directory of Open Access Journals (DOAJ)[2]. However, open access was not included in the final conclusions because the obtained results seemed inconsistent with previous work (Eysenbach, 2006) and because the data from DOAJ may be inaccurate. Regarding the field type factor, the same OECD classification used in Gazni and Thelwall (2015) was used in this study. Each publication in WoS has a subject area that was mapped into one of the OECD sub-fields[3]. The sub-fields were categorized into six fields including Natural Sciences, Engineering & Technology, Medical & Health Sciences, Agricultural Sciences, Social Sciences, and Humanities using OECD mapping instruction[4]. Then, the six fields were mapped into three broad fields resulting in all publications having been categorized into the three fields: Social Sciences & Humanities; Medical & Natural Sciences; and Engineering & Technology. To measure the field size, each article was assigned to a sub-category. The OECD sub-categories given to the articles were used for this purpose. Field size was calculated by the number of publications, number of authors, number of journals, number of institutions, and number of countries in the sub-category. These measures were highly correlated and so only field size (in terms of the number of publications) was used in the models. Regarding the abstract readability, the Flesch Reading Ease Score was used as it is the most popular measure of text readability. The Flesch Score ranges between 0 and 100, where 0 indicates a text that is the most difficult to read and 100 represents the easiest text to read. Table 1 summarizes how each of the variables were measured.

---

[2] https://doaj.org/

*Statistical modeling:* Given that the dependent variables of this study are count data (citation and altmetric counts), count regression models are the most appropriate models. The basic count models are Poisson and Negative Binomial (NB) models. Since the data in this study is over dispersed, the Poisson model, in which the mean and the variance are assumed to be equal (Cameron & Trivedi, 1998), is not appropriate, whereas the NB model addresses this type of data. Furthermore, the data has more zeros than are accounted for in the NB distribution requiring a count model that can deal with excess zeroes. A negative binomial-logit hurdle model is the best fit for the data; this model creates a scenario in which the positive counts follow a Poisson or NB distribution after passing a hurdle in order to gain positive counts. The model has two parts: a negative binomial part that models the positive non-zero observations and a binary (or logit) part that models the zero observations. Hence, the significant factors of both positive counts and zero counts of dependent variables can be determined through the two parts of the model. The hurdle model is also preferred since it simultaneously assesses a number of factors with citation and altmetric counts rather than the simpler regression models that separately test factors, which may generate inappropriate models.

Table 1. Factors and measures

| Factors | Measures |
| --- | --- |
| **Journal impact factor (JIF)** | Journal impact factor retrieved from JCR for the publishing journal in the WoS 'SO' field for the publication. |
| **Individual collaboration (No. Authors)** | Number of authors listed in the WoS 'AU' field for the publication. |
| **International collaboration (No. Countries)** | Number of different country names listed in the WoS 'C1' field for the publication. |
| **Institution prestige** | Maximum Mean Normalized Citation Score (MNCS) of different institution names listed in the WoS 'C1' field for the publication. |
| **Country prestige** | Maximum Mean Normalized Citation Score (MNCS) of different country names listed in the WoS 'C1' field for the article. |
| **Research funding** | Funded (1) if there is an entry in the WoS 'FU' field for the article; Unfunded (0) if there is no entry in the WoS 'FU' field for the article. |

---

[3] http://incites.isiknowledge.com/common/help/h_field_category_oecd_wos.html
[4] http://incites.isiknowledge.com/common/help/h_field_category_oecd.html

| | |
|---|---|
| **Abstract readability (Abs. Readability)** | Flesch readability score of the abstract in the WoS 'AB' field for the article. |
| **Abstract length (Abs. Length)** | Number of words in the abstract in the WoS 'AB' field for the article. |
| **Title length** | Number of words in the title in the WoS 'TI' field for the article. |
| **Number of cited references (No. references)** | Number of references listed in the WoS 'CR' field for the article. |
| **Field size** | Number of publications in the related sub-field. |
| **Field type** | OECD field of paper: Social Sciences & Humanities (1); Engineering & Technology (2); Medical & Natural Sciences (3) |

## Results

The results are presented in six sections: citation counts, Mendeley readers, Twitter posts, Facebook posts, and blog and news posts, respectively. In each section, the results of hurdle models are discussed in two parts: the first part presents the results of the negative binomial component of the model demonstrating the factors that associate with increased or decreased citation or altmetric counts; the second part presents the results of the logit component demonstrating the factors that associate with zero citation or altmetric counts.

*1. Citation counts:* The results of the hurdle model for citation counts are presented in Table 2. The negative binomial component of the model shows that all factors are significant factors of citations except for field size. Journal impact factor increases the number of citations by 7.7%. International collaboration also associates with increased citation counts and a unit change in the factor contributes to 6.2% increased citations, whereas individual collaboration contributes to decreased citation counts. Institution and country prestige also associate with increased tweets and a unit change in each factor associates with 2.5% and 0.1% increase, respectively. Research funding significantly associates with increased citation counts and a unit increase in the factor associates with 40.3% increase in the number of citations. While articles with longer abstracts receive 0.1% more citations, articles with easier abstracts to read potentially receive 0.7% less citations than the articles with more difficult abstracts. Articles with shorter titles by one word receive 1.6% more citations and articles with one more reference in their reference lists receive 0.5% more citations. Field type significantly associates with citation counts demonstrating that Medical & Natural Sciences research articles are 51.2% more likely cited than articles in Engineering & Technology and Social Sciences & Humanities.

The logit component of the model establishes that international collaboration, abstract readability, abstract length, title length, and field size are insignificant, while journal impact factor, institution prestige, country prestige, research funding, and number of references significantly associate with decreased zero citations. Journal impact factor, institution prestige, country prestige, research funding, number of references, and field type decrease the number of zero citations by 15%, 0.8%, 0.1%, 42.7%, 2%, and 60.6% respectively. Individual collaboration significantly associates with increased zero citation counts and a unit change in the factor increases zero counts by 7.3%.

Table 2. The results of hurdle models for citation counts and Mendeley readership

| | Citations | | | Mendeley | | |
|---|---|---|---|---|---|---|
| Logit model | Coef. | Exp(Coef.) | P>z | Coef. | Exp(Coef.) | P>z |
| JIF | 0.14 | 1.15 | 0.000 | 0.001 | 1.001 | 0.002 |
| No. Authors | -0.006 | 0.994 | 0.007 | -0.040 | 0.961 | 0.000 |
| No. Countries       | 0.015  | 1.015      | 0.824 | 0.007  | 1.007      | 0.000 |
| Institution prestige| 0.008  | 1.008      | 0.012 | 0.037  | 1.038      | 0.002 |
| Country prestige    | 0.001  | 1.001      | 0.010 | 0.024  | 1.024      | 0.000 |
| Research funding    | 1.232  | 3.427      | 0.000 | -0.055 | 0.946      | 0.864 |
| Abs. Readability    | 0.004  | 1.004      | 0.505 | 0.190  | 1.209      | 0.000 |
| Abs. Length         | 0.001  | 1.001      | 0.332 | 0.029  | 1.029      | 0.019 |
| Title length        | 0.001  | 1.001      | 0.928 | -0.031 | 0.969      | 0.308 |
| No. References      | 0.02   | 1.02       | 0.000 | 0.015  | 1.015      | 0.017 |
| Field size          | 0.000  | 1.000      | 0.237 | 0.000  | 1.000      | 0.233 |
| Field type          | 0.474  | 1.606      | 0.000 | -0.101 | 0.904      | 0.392 |
| NB model            | Coef.  | Exp(Coef.) | P>z   | Coef.  | Exp(Coef.) | P>z   |
| JIF                 | 0.075  | 1.077      | 0.000 | 0.001  | 1.001      | 0.000 |
| No. Authors         | -0.001 | 0.999      | 0.009 | -0.019 | 0.981      | 0.000 |
| No. Countries       | 0.06   | 1.062      | 0.001 | 0.004  | 1.004      | 0.000 |
| Institution prestige| 0.025  | 1.025      | 0.000 | 0.030  | 1.031      | 0.000 |
| Country prestige    | 0.001  | 1.001      | 0.000 | 0.013  | 1.014      | 0.000 |
| Research funding    | 0.338  | 1.403      | 0.000 | 0.250  | 1.283      | 0.000 |
| Abs. Readability    | -0.007 | 0.993      | 0.01  | 0.023  | 1.023      | 0.000 |
| Abs. Length         | 0.001  | 1.001      | 0.002 | -0.001 | 0.999      | 0.000 |
| Title length        | -0.016 | 0.984      | 0.006 | -0.047 | 0.954      | 0.000 |
| No. References      | 0.005  | 1.005      | 0.000 | 0.009  | 1.009      | 0.000 |
| Field size          | 0.000  | 1.000      | 0.5   | 0.000  | 1.000      | 0.171 |
| Field type          | 0.413  | 1.512      | 0.000 | -0.173 | 0.841      | 0.000 |

*2. Mendeley readership:* The results of the negative binomial model demonstrate that all factors significantly associate with increased readers except for field size. Journal impact factor associates with an increased number of readers and a unit increase in the factor increases the number of readers by 0.1%. While international collaboration increases the probability of more readership by 0.4%, individual collaboration decreases this chance by 1.9%. Articles from more prestigious institutions and countries are more often added to Mendeley and likely receive 3.1% and 1.4% more readers than articles from less prestigious institutions and countries. Funded research will likely have 28.3% additional readers than unfunded research. Articles with easier abstracts have 2.3% additional readers than the articles with more difficult abstracts, but articles with longer abstracts receive 0.1% less readers. Articles with longer titles also have 4.6% less readers, but articles with a longer list of references have 0.9% additional readers. Finally, there is a negative association between field type and number of readers indicating that articles in Social Sciences & Humanities have the largest number of readers; Medical & Natural Sciences and Engineering & Technology are predicted to have 19.7% less readers than that of articles from Social Sciences & Humanities.

When examining the results of the logit model, the following factors do not significantly associate with zero readership rates: research funding, title length, field size, and field type. Journal impact factor, international collaboration, institution and country prestige, abstract readability, abstract length, and number of references significantly associate with decreased zero citations, whereas individual collaboration contributes to increased zero citations (Table 2).

*3. Tweets*: Regarding the negative binomial model, all factors except for institution prestige, country prestige, and title length significantly associate with number of tweets. Journal impact factor significantly associates with an increased number of tweets and a unit change in the factor increases the number of tweets by 11.1%. Individual and international collaborations also associate with increased tweets, although individual collaboration demonstrates a weak association (articles with one additional author potentially receive 0.1% more tweets). International collaboration strongly associates with increased tweets and one additional country likely contributes to a 9.5% increase in the number of tweets. Research funding has the strongest association with number of tweets; funded research receives 24.6% more tweets than unfunded research. Articles with longer and easier abstracts to read receive 0.5% more tweets. Furthermore, articles with a longer list of references likely receive 0.2% more tweets. Articles from bigger subject fields receive 0.1% more tweets. Additionally, articles in Medical & Natural Sciences and Engineering & Technology receive 14.7% less tweets than that of Social Sciences & Humanities.

The logit model shows that only journal impact factor and individual collaboration significantly associate with decreased zero tweets, while all other factors are insignificant factors of zero number of tweets (Table 3).

Table 3. The results of hurdle models for Twitter and Facebook

| | Twitter | | | Facebook | | |
|---|---|---|---|---|---|---|
| Logit model | Coef. | Exp(Coef.) | P>z | Coef. | Exp(Coef.) | P>z |
| JIF | 0.079 | 1.082 | 0.011 | 0.064 | 1.066 | 0.000 |
| No. Authors | 0.001 | 1.001 | 0.007 | -0.002 | 0.998 | 0.003 |
| No. Countries | -0.065 | 0.937 | 0.423 | -0.037 | 0.964 | 0.303 |
| Institution prestige | 0.000 | 1.000 | 0.948 | -0.003 | 0.997 | 0.378 |
| Country prestige | 0.000 | 1.000 | 0.928 | 0.000 | 1.000 | 0.600 |
| Research funding | 0.069 | 1.071 | 0.564 | 0.248 | 1.282 | 0.042 |
| Abs. Readability | -0.004 | 0.996 | 0.305 | -0.002 | 0.998 | 0.603 |
| Abs. Length | 0.001 | 1.001 | 0.338 | 0.002 | 1.002 | 0.002 |
| Title length | -0.003 | 0.997 | 0.851 | -0.007 | 0.993 | 0.579 |
| No. References | -0.001 | 0.999 | 0.429 | 0.002 | 1.002 | 0.010 |
| Field size | 0.000 | 1.000 | 0.101 | 0.000 | 1.000 | 0.406 |
| Field type | 0.079 | 1.082 | 0.363 | -0.107 | 0.898 | 0.097 |
| NB model | Coef. | Exp(Coef.) | P>z | Coef. | Exp(Coef.) | P>z |
| JIF | 0.106 | 1.111 | 0.000 | 0.068 | 1.070 | 0.000 |
| No. Authors | 0.001 | 1.001 | 0.002 | 0.000 | 1.000 | 0.545 |
| No. Countries | 0.091 | 1.095 | 0.023 | 0.133 | 1.142 | 0.003 |
| Institution prestige | -0.001 | 0.999 | 0.677 | 0.006 | 1.006 | 0.408 |
| Country prestige | 0.000 | 1.000 | 0.135 | 0.000 | 1.000 | 0.246 |
| Research funding | 0.220 | 1.246 | 0.007 | 0.621 | 1.862 | 0.005 |
| Abs. Readability | 0.005 | 1.005 | 0.045 | -0.017 | 0.983 | 0.201 |
| Abs. Length | 0.005 | 1.005 | 0.000 | 0.003 | 1.003 | 0.102 |
| Title length | -0.015 | 0.985 | 0.185 | -0.059 | 0.942 | 0.019 |
| No. References | 0.002 | 1.002 | 0.016 | 0.002 | 1.002 | 0.184 |
| Field size | 0.001 | 1.001 | 0.013 | 0.000 | 1.000 | 0.079 |
| Field type | -0.159 | 0.853 | 0.026 | -0.272 | 0.762 | 0.048 |

*4. Facebook posts*: With respect to the negative binomial model, individual collaboration, institution prestige, country prestige, abstract readability and length, number of references, and field size are insignificant factors of number of Facebook posts. However, journal impact factor, international collaboration, and research funding are significantly positively associated with increased posts. A unit increase in the journal impact factor, international collaboration, and research funding increases the number of posts by 7%, 14.2%, and 86.2%, respectively. Title length associates with a decreased number of Facebook posts indicating that articles with longer titles receive 5.8% less posts than the articles with shorter titles. Field type also associates with decreased posts demonstrating that Medical & Natural Sciences and Engineering & Technology research likely receives 28.3% less posts than Social Sciences & Humanities research.

With respect to the logit model, only journal impact factor, research funding, abstract length, and number of references significantly associate with decreased zero posts. Individual collaboration significantly associates with increased zero posts, but all other factors are insignificant factors of zero Facebook posts (Table 3).

*5. Blog posts*: The negative binomial model results finds that research funding, abstract readability, abstract length, and number of references are not significant factors for blog posts, but all other factors significantly associate with the number of posts. Journal impact factor significantly associates with increased blog posts and a unit increase in the factor increases the number of posts by 7%. Individual and international collaborations also significantly associate with increased blog posts; one additional author and one additional country in the paper contributes to 0.1% and 15.3% increase in the number of posts, respectively. Institution and country prestige associates with 1.7% and 0.1% increase in the number of posts, respectively. Field size significantly associates with increased blog posts, whereas field type associates with decreased number of posts. Articles from bigger fields have been more blogged than the articles from smaller domains. Articles from Medical & Natural Sciences and Engineering & Technology likely have 37.2% less blog posts than that of Social Sciences & Humanities. Title length significantly associates with decreased posts indicating that articles with shorter titles have 7.2% more blog posts than articles with longer titles.

With respect to the logit model, abstract readability, abstract length, and number of references are insignificant factors of zero posts, while the other factors significantly associate with zero posts. Among the significant factors, only title length and field type significantly associates with increased zero posts. Journal impact factor, number of authors, number of countries, institution prestige, country prestige, research funding, and field size significantly associate with decreasing zero blog mentions (Table 4).

Table 4. The results of hurdle models for blog and news

| | Blog | | | News | | |
|---|---|---|---|---|---|---|
| Logit model | Coef. | Exp(Coef.) | P>z | Coef. | Exp(Coef.) | P>z |
| JIF | 0.092 | 1.096 | 0.000 | 0.088 | 1.092 | 0.000 |
| No. Authors | 0.001 | 1.001 | 0.000 | -0.002 | 0.998 | 0.014 |
| No. Countries | 0.086 | 1.090 | 0.043 | 0.019 | 1.019 | 0.676 |
| Institution prestige | 0.018 | 1.018 | 0.000 | 0.014 | 1.015 | 0.000 |
| Country prestige | 0.001 | 1.001 | 0.000 | 0.001 | 1.001 | 0.000 |
| Research funding | 0.689 | 1.991 | 0.003 | 0.337 | 1.401 | 0.029 |
| Abs. Readability | -0.006 | 0.995 | 0.493 | -0.019 | 0.981 | 0.001 |
| Abs. Length | 0.000 | 1.000 | 0.689 | 0.000 | 1.000 | 0.831 |
| Title length | -0.049 | 0.952 | 0.000 | -0.057 | 0.944 | 0.000 |
| No. References | -0.004 | 0.996 | 0.092 | -0.006 | 0.994 | 0.028 |
| Field size | 0.001 | 1.001 | 0.030 | 0.001 | 1.001 | 0.249 |
| Field type | -0.245 | 0.783 | 0.007 | -0.341 | 0.711 | 0.001 |
| NB model | Coef. | Exp(Coef.) | P>z | Coef. | Exp(Coef.) | P>z |
| JIF | 0.067 | 1.070 | 0.000 | 0.043 | 1.044 | 0.000 |
| No. Authors | 0.001 | 1.001 | 0.022 | 0.007 | 1.007 | 0.026 |
| No. Countries | 0.142 | 1.153 | 0.001 | 0.136 | 1.146 | 0.022 |
| Institution prestige | 0.017 | 1.017 | 0.000 | 0.021 | 1.022 | 0.000 |
| Country prestige | 0.001 | 1.001 | 0.000 | 0.001 | 1.001 | 0.000 |
| Research funding | 0.089 | 1.093 | 0.799 | -0.083 | 0.921 | 0.791 |
| Abs. Readability | -0.001 | 0.999 | 0.926 | -0.029 | 0.971 | 0.009 |
| Abs. Length | -0.005 | 0.995 | 0.25 | 0.000 | 1.000 | 0.983 |
| Title length | -0.074 | 0.928 | 0.000 | -0.020 | 0.980 | 0.460 |
| No. References | 0.003 | 1.003 | 0.660 | -0.006 | 0.994 | 0.053 |
| Field size | 0.001 | 1.001 | 0.004 | 0.000 | 1.000 | 0.766 |
| Field type | -0.464 | 0.628 | 0.032 | -0.516 | 0.597 | 0.046 |

*6. News posts*: Results from the negative binomial model found that all factors except for research funding, abstract length, title length, number of references, and field size, are significantly associated with increased news posts. Journal impact factor associates with increased posts and a unit increase in the factor contributes to 4.4% increase in the number of news posts. Individual and international collaborations also associate with increased posts and a unit increase in each factor increases the number of posts by 0.7% and 14.6%, respectively. Institution and country prestige associates with increased news posts and a unit increase in each factor contributes to 2.2% and 0.1% increase in the number of posts, respectively. Abstract readability associates with decreased news posts showing that articles with easier abstracts are less mentioned in news. Articles from Social Sciences & Humanities are mentioned more in news posts than from Engineering & Technology and Medicine & Natural Sciences.

With respect to the logit model, journal impact factor, institution prestige, country prestige, and research funding significantly associate with decreased zero posts, while individual collaboration, abstract readability, title length, number of references, and field type significantly associate with increased zero posts (Table 4).

In the following discussion section, the results are presented factor by factor and followed by an examination of the results as framed by the research questions.

## Discussion

Journal impact: As widely confirmed in previous studies (Didegah, 2014), journal impact is the most important determinant of citations. Similarly, it was found to be an important factor for both citations and altmetric events in the current study. In the case of citations, it is perceived that top journals contain higher quality content and thus they are cited more. Social media users also tend to choose higher quality content as high impact journals are more read on Mendeley, more tweeted, more posted on Facebook, and more mentioned in blog and news posts. It is interesting to note that the journal impact factor effect is even higher on Twitter posts than on citations showing that high impact journals are also popular on social media platforms; however, it is important to know that majority of journals particularly top journals such as Nature, Science, PLoS One, etc. have Twitter accounts and tweet their recent articles regularly. Individual and international collaboration: This factor is significantly associated with increased numbers of tweets, and blog and news posts, demonstrating that the more authors a paper contains, the higher the number of tweets and blog and news posts the paper receives. More collaborators are assumed to increase visibility as the more authors in a paper, the more people who know them and look their articles up, share them or comment on them. The results for citation counts contradict the results of prior research, i.e. the positive association between the number of authors and citation counts (Chen, 2012; Franceschet & Costantini, 2010; Gazni & Didegah, 2010; Persson, 2010). A recent study examined this factor for altmetric events (i.e. blog posts, Twitter users, public Facebook shares, Google+, and news and mainstream media) and found that the number of authors was an important factor for all altmetric events examined (Haustein, Costas, & Larivière, 2015), which confirms the results of this work for the number of tweets, blog posts and news posts, but contradicts the results for Facebook posts. This contradiction may be due to several reasons including different data samples—for example, this work investigates research articles by Finnish researchers over three years, while the earlier work (Haustein, Costas & Lariviére, 2015) examined all Web of Science documents from 2012—and also because the earlier work used simpler statistical tests to model the factors rather than simultaneously modelling the factors, as was done in this work. Moreover, the limitation with Facebook data that only public posts are accessible may have also affected the results. The international collaboration is an important factor for both citation and altmetric counts. The more countries collaborating on a paper, the higher the number of citations and also all altmetric counts to the paper. The international collaboration has been found to be significant for increased citations in the majority of previous studies (Sin, 2011; Persson, 2010). This significant association with altmetric counts also concurs with the results of Haustein, Costas, and Larivière (2015). Researchers are active on social media such as Twitter and tweet about research outputs (Tsou, Bowman, Ghazinejad, & Sugimoto, 2015). Therefore, having an international network can also have an influence on the author connection on social media and boost their online network towards more international connections. And more online connections will definitely increase the author visibility and their articles visibility consequently. Institution and country prestige: Similar to Didegah (2014), the results indicate that papers published by top ranked institutions and countries receive more citations. The finding s show that scholars have a propensity towards citing and reading publications from prestigious institutions and countries, while the institution is not important for tweeters or Facebook users. The similarity between citations and Mendeley readership is (most likely) due to similar types of users, as previous work has found that Mendeley users are primarily scholars (Mohammadi, Thelwall, Haustein, & Lariviére, 2015); however, in a study of tweeters who tweeted about articles from four top journals (Nature, Science, PLOS ONE, and PNAS), nearly half of the tweeters were part of the scholarly community (Tsou, Bowman, Ghazinejad, & Sugimoto, 2015). The difference between these platforms could, however, be explained by other reasons such as that Twitter and Facebook are mainly used to consume or disseminate research that discusses current trending topics and important events or the research that is freely accessible on the web regardless of which institution was associated with the research, while scholars may cite publications from top institutions in order to lend more attention to their own work or to persuade journal editors or other authors about the high quality of their research. Research funding: The results show that funding has a very strong impact on citation counts and that funded articles receive 40.3% more citations than unfunded articles. This has also been claimed in previous studies (Zhao, 2010), although the association between citation impact and research funding varies across different subject fields (Didegah, 2014). The factor is also important for the number of Mendeley readers, tweets, and Facebook posts, but it is not an important factor for the number of blog posts and news posts. Research funding supports scientists and paves the way for creative and high-quality research, especially in experiment-based research fields. Insufficient funds may lead to shortcomings in the research as researchers may not be able to do high quality work due to a lack of equipment and materials. Because scholars may assess funded research as being of a higher quality than unfunded research, they may then favor saving funded research to their Mendeley library. Funded research is also more mentioned on Twitter and Facebook than unfunded research, which could be caused by requirements placed on the researchers by the funding agencies that charge them with providing evidence of social impact. These requirements may cause the researchers to be more active on Twitter and Facebook with regards to promoting their own research and the research of others. Abstract readability: Articles with more difficult abstracts to read receive more citations, and are more reported in the news, while articles with easier abstracts to read are added more to Mendeley and tweeted more often on Twitter. Previous studies also demonstrated that articles with more difficult abstracts are cited more (Didegah, 2014), which may be due to the fact that scholars are experts in their own fields and have prior knowledge of their complex terminologies. Thus, a text may be considered difficult for non-experts as indicated by Flesch scores, but it may be easy to understand for the experts. This notion of expertise may also apply to news posts, as some of the news posts are written by journalists who may specialize in a specific topic and thus have a better understanding of the specialized vocabulary. While it has been found that many of those tweeting about science are scholars (Tsou, Bowman, Ghazinejad, & Sugimoto, 2015), those scholars who do tweet may be choosing to tweet about articles with abstracts that are easier to read because they know that many of their own followers are members of the general public. Mendeley readers also tend to add articles with easier abstracts to their library indicating that Mendeley readers may have a general readership behavior that does not limit them to exclusively read articles from their own field. Abstract readability is not an important factor for Facebook and blog posts. Abstract and title length: The longer the article abstract, the more citations and tweets. One reason explaining why an article with a longer abstract may have more citations and social media mentions is that an extensive abstract is a more complete representation of a paper, providing citers and tweeters with more details and enabling them to make a decision about the usefulness of the work. However, it seems Mendeley readers, Facebookers, bloggers and news providers do not mind how long an article abstract is and they take some other important features of the article into accoun t. Moreover, the shorter the article title the more it is cited, which has also been widely confirmed in previous studies that shorter titles have a citation advantage (Didegah, 2014; Ayres & Vars, 1999). Shorter titles are more attractive in Mendeley also and it may be that longer titles cannot easily draw a reader's attention to the main message of the article, whereas a shorter title may increase a reader's ability to filter relevant articles from the plethora of available academic documents. Number of references: References in an article attract other researchers to the article, researchers who are searching for relevant studies through citations. Therefore, more references in the article increase the chance of more attention to the article. Higher online and offline attention to works with more references is expected since the more the references, the more comprehensive the paper is expected; and that reference-based searching in databases increases the chance of more visibility. In the traditional citation context, there is a hypothesis that authors tend to cite the works of their ex-citers (Webster, Jonason, & Schember, 2009) showing that more one cites, more he/she receives citations in return. However, this hypothesis may not be applied to the online platforms such as Mendeley and Twitter and some other reasons may influence such as that more references in the article act as more channels to find that article more easily and quickly among so many other relevant works. Field size and type: Field size is not an important factor for citation counts, nor is it for altmetric scores. How big a field is in terms of authors or publications influence its visibility and impact. Articles from smaller fields are assumed to be less visible than the articles from broader fields (King, 1987). The field size affects some other factors such as the number of references which is an important factor for increasing visibility and articles from broad fields are found to have longer list of references. Field type is an important positive factor for citation counts, but its association with all altmetric events is negative. This shows that articles from Medical & Natural Sciences are more cited than those from Engineering & Technology, and articles from Engineering & Technology are also cited more than those from Social Sciences & Humanities. It is, however, the opposite for altmetric events as Social Sciences & Humanities articles have more mentions on altmetric platforms than articles from Engineering & Natural sciences. This could be due to the readability of Social Sciences & Humanities research (easier to read and comprehend than Engineering, Medicine & Natural Sciences research). To answer the first research question, the results indicate that except for field size, all other factors are driving both citations and all, or some, of the altmetric events. Journal impact factor and international collaboration are the two factors that significantly associate with increased citation counts and all altmetric scores. Therefore, it seems that some of the citation factors are also important for altmetric counts. As clearly shown in Table 5, citation and Mendeley readership are only behaving differently based on three factors: abstract readability, abstract length, and field type. The factors driving other altmetric events do behave much more differently from the way they are driving citations. To answer the second research question, the data supports the conclusion that the altmetric events differ from each other in terms of specific factors. For instance, while institution prestige and country prestige do not matter for Twitter posts and Facebook posts, they significantly associate with increased blog and news posts. As shown in Table 5, blog and news posts have similar associations across nine factors, whereas Twitter and Facebook have similarity in association across seven factors.

Table 5. Factors significance across different platforms

| Factor | Citation | Mendeley | Twitter | Facebook | Blog | News |
|---|---|---|---|---|---|---|
| JIF | 7.7 | 0.1 | 11.1 | 7 | 7 | 4.4 |
| No. Authors | -0.1 | -1.9 | 0.1 | insig | 0.1 | 0.7 |
| No. Countries | 6.2 | 0.4 | 9.5 | 14.2 | 15.3 | 14.6 |
| Institution prestige | 2.5 | 3.1 | insig | insig | 1.7 | 2.2 |
| Country prestige | 0.1 | 1.4 | insig | insig | 0.1 | 0.1 |
| Research funding | 40.3 | 28.3 | 24.6 | 86.2 | insig | insig |
| Abs. readability | -0.7 | 2.29 | 4.5 | insig | insig | -2.9 |
| Abs. length | 0.1 | -0.1 | 0.5 | insig | insig | insig |
| Title length | -1.6 | -4.6 | insig | -5.8 | -7.2 | insig |
| No. References | 0.5 | 0.9 | 0.2 | insig | insig | insig |
| Field size | insig | insig | 0.1 | insig | 0.1 | insig |
| Field type | 51.2 | -15.9 | -14.7 | -23.8 | -37.1 | -40.3 |

Negative significant                                    Positive significant

# Conclusion

This research provides an increased understanding of factors that may affect the ways in which scientific work receives online attention. While the results found in this work partly differ from the research by Haustein, Costas, and Larivère (2015) (most likely due to the differences in statistical tests used, the difference in samples, and differences in time frames), what is important is that these results point to the uncertainty that comes from studying altmetric events that are being captured from constantly changing ecosystems with a large, (mostly) invisible user base. As researchers continue to develop theoretical and methodological models to study this context, the different results found in this early stage of research should be made clear. What is apparent is that more research is needed using different models, theories, and populations to study these phenomena. The generalisation of the current findings to other contexts and other countries is dependent on several factors; the countries should have fairly similar overall research profiles, science communication in social media should be at the same level, and publishing practices should be similar. In this study Finland was chosen as the object to study and no comparisons to other countries were thus made. Future research could focus on disciplinary differences, in which case the results might be more generalisable. The findings from this study can contribute to the continued development of theoretical models and methodological developments associated with capturing, interpreting, and understanding altmetric events. This work can also aid research policy makers with identifying important factors driving altmetric events.

# Funding

This research was financed by The Finnish Ministry of Education and Culture's Open Science and Research Initiative 2014–2017 (funding number: OKM/33/524/2015).

# Acknowledgement

The authors wish to thank Vincent Larivière and Stefanie Haustein for their comments and suggestions during preparing this manuscript.